# Comparative study of wall shear stress at the ascending aorta for different mechanical heart valve prostheses

Qianhui Li and Christoph H. Bruecker

School of Mathematics, Computer Science and Engineering. City, University of London, Northampton square, London, EC1V 0HB, UK

**Abstract**

An experimental study is reported which investigates the wall shear stress (WSS) distribution in a transparent model of the human aorta comparing a bileaflet mechanical heart valve (BMHV) with a trileaflet mechanical heart valve (TMHV) in physiological pulsatile flow. Elastic micro-pillar WSS sensors, calibrated by micro-Particle-Image-Velocimetry measurement, are applied to the wall along the ascending aorta (AAo). Peak WSS values are observed almost twice in BMHV compared to TMHV. Flow field analyses illuminate that these peaks are linked to the jet-like flows generated in the valves interacting with the aortic wall. Not only the magnitude but also the impact regions are specific for the different valve designs. The side-orifice jets generated by BMHV travel along the aortic wall in the AAo and cause a whole range impact, while the jets generated by TMHV impact further downstream in the AAo generating less severe WSS.

## 1. Introduction

The heart valves of the left ventricle of the human heart are subject to the highest mechanical loads of all valves and therefore are often the first to deteriorate in elderly people[1]. The mitral valve and the aortic valve control the unidirectional flow of oxygenated blood from the lung towards the arterial system via the left ventricle muscle activity. Diseases of both valves lead to a morbidity rate of 44% and 34% respectively[2, 3] if not treated successfully by either repair or replacement with artificial prostheses. Compared to the common coronary heart disease, although the pervasiveness of the heart valve disease (HVD) is small, the effect of HVD is disproportionately huge, because of the need of long-term follow-up, and expensive treatment costs and investigation[4]. Concomitantly, it is reported that the number of patients in need of heart valve replacement is almost 290,000 in 2003, which is estimated to be triple by 2050[5]. However, the replacement of artificial heart valves has been proved to cause a wide variety of valve-related complications, causing sometimes severe disability or even death[6]. One main reason is that hemodynamics induced by artificial heart valves is different from the natural valve. There is a huge requirement to better qualify hemodynamics performance of artificial heart valves and make improvement in the design. Fluid mechanics of artificial heart valve has attracted extensive attention in the past, for review papers see Sotiropoulos et al. [7] and Yoganathan et al. [3]. The additional jets generated by artificial heart valves usually generate much higher velocity gradients than the natural valve[8]. In addition to the complex core flow, the interaction between the vessel walls and the jets is also of great importance, since high-velocity gradient, i.e. high magnitude of wall shear stress (WSS), on the impacting region is formed. WSS provides a connection between the fluid mechanics of blood flow and the biology affecting many cardiovascular diseases. Oscillating shear forces plus large WSS gradients probably promote the development of intimal hyperplasia (IH).[9] Moreover, a high WSS could lead to stress-related damage of the vessel's inner walls [10, 11]. Besides, aortic aneurysm rupture is believed to occur when the mechanical stress acting on the wall, composed of pressure forces and the WSS, exceeds the strength of the wall tissue[12]. It has also been reported that vessel segments under the impact of low WSS or highly oscillatory WSS are at the highest risk to develop atherosclerosis. Therefore, it is expected that WSS mapping could contribute to the multidisciplinary and multifactorial approach to detect early atherosclerosis[13].

It is obvious that there is an urgent requirement to qualify the valve prosthesis not only from a hydraulics viewpoint but to include also the connecting vessel walls into the design and qualification process. The stresses acting on the walls are composed of the normal stresses (pressure) and the tangential stresses (WSS), both being important to estimate the load of the endothelial cells at the walls. Although it has been reported that the oscillation, low or high magnitude of the WSS can lead to severe heart diseases,[10, 11] to our knowledge, most of the previous studies[1, 8, 14] on the qualification of the artificial valves mainly focus on the flow field in the vicinity of artificial valves in simplified straight vessel models. Only very little attention is paid to the effects of the implanted artificial heart valves on WSS magnitude and distribution along the aortic wall[15]. As for measurement methods, there are quite a lot of limitations to perform measurements in a natural condition, as the test section are organs in vivo and the fluid is blood. The limited spatial resolution of the in-vivo measurements such as magnetic resonance imaging (MRI)[16] leads to huge challenge to measure WSS accurately. Also, it is high-risk to qualify a newly designed artificial heart valve by implanting it into a human body. The in-vitro measurement such as Particle Image Velocimetry (PIV) still require optical access. It is also difficult to apply direct measurements by using MEMS or other WSS sensors to such studies. In our study, we applied flexible micro-pillar sensors, for the first time, to imaging of the wall shear stress along the AAo. The micro-pillar sensors have a broad application prospect, they could be even applied to a non-transparent fluid such as blood if markers are attached to the sensor tips. Calibration is done herein with highly resolved PIV measurements in a transparent flow rig.

**2. Materials and Methods**

**Heart valve prosthesis**

Artificial heart valves are divided into two major types, mechanical heart valves (MHV) and bioprosthetic heart valves. Until recently, heart valve replacement was commonly performed using MHV, which offer a life-long durability[17]. We herein focus on the comparative study of MHV, including SJM Regent BMHV (St. Jude Medical Inc., Minnesota, U.S.) and Triflo TMHV (Triflo Medical Inc., Neuchâtel, Switzerland) with the same diameter $D_A$ = 25 mm (see Fig. 1). SJM Regent valve is a wildly used MHV and therefore taken as a control valve in our comparative study. Nonetheless, clinical reports indicate that conventional MHVs are still unable to avoid the problems of cardiovascular diseases and complications on the long-term anticoagulation[6]. Triflo TMHV consists of three leaflets that open to form a central orifice for the blood to flow through and three side orifices jets. The design of TMHV more closely resembles the native aortic valves and is assumed to have a better hemodynamic performance.

An open circular orifice nozzle is also used to perform in-situ micro-PIV measurements for the calibration of the micro-pillar sensors.

**Experimental setup**

The experimental setup is described in detail in our previous studies[18, 19] and is shortly summarized herein. The geometry data of the model aorta are taken from Vasava et al.[20], who simplified the geometry derived from the three-dimensional reconstruction of a series of two-dimensional slices obtained in vivo using CAT scan imaging on a human aorta by Shahcheraghi et al. [21] (see table 1). Although the geometrical dimensions of human aorta may slightly vary with age and sex, the comparative effects observed between the bileaflet mechanical heart valve (BMHV) and the trileaflet mechanical heart valve (TMHV) should represent the relative differences in a realistic manner. The diameter of the aorta is $D_A$=25mm and the MHVs are selected to properly fit the size. The 180° curved bend representing the aortic arch with sinuses of Valsalva (SOV) has a planar symmetry plane (crossing the left sinus in the middle), which is useful for our model fabrication and sensor integration. The transparent model of the aorta is the same as used in [18]. It consists of two symmetric halves made out of silicone (Elastosil RT 601™, refractive index n ≈ 1.4095) which are later assembled together as shown in Fig. 2(b). The sensors are integrated into the model with a thin sheet (thickness of 100 µm), which is clamped between both halves. This sheet is made from Polydimethylsiloxane (PDMS) (density $\rho_{PDMS}$ ≈ 1030 kg·m$^{-3}$, Young's modulus $E \approx 1.24$ MPa) and carries at its inner contour twelve flexible micro-pillars that extend out of the contour (see Fig. 2(c)). The curved contour of the sheet is adapted to the geometry of the inner wall of the model such that it is flush with the wall with the flexible micro-pillars protruding into the inner of the flow. There, the micro-pillars are subject to drag forces proportional to the WSS, which causes their bending relative to the situation when the liquid is at rest.

| Artery properties | Dimension (mm) |
|---|---|
| (1) Diameter of sinus bulb | 21 |
| (2) Lumen diameter of ascending and descending aorta | 25 |
| (3) Lumen diameter of brachiocephalic artery (BA) | 8.8 |
| (4) Lumen diameter of left common carotid artery (LCA) | 8.5 |
| (5) Lumen diameter of left subclavian artery (LSA) | 9.9 |

**Table 1.** Main geometry data of the aorta model, taken from Vasava et al.[20], see also [19].

The transparent aorta is then placed at the bottom of a transparent container (width/depth: 190 mm, height: 500 mm), which is filled with a glycerine-water solution (58/42 % by mass, density $\rho$ = 1140 kg m$^{-3}$, dynamic viscosity $\mu$= 5 mPa·s at the temperature of 38°C, refractive index 1.4095). This mixture has the same refractive index as the silicone model and therefore ensures undisturbed optical access into the inner of the aorta. A picture of the view onto a millimetre paper at the back of the model demonstrates the good match of both refractive index, see Fig. 2(d). Small tracer particles (fluospheres, mean diameter 30micron, Dantec Dynamics) are added to the liquid for the PIV recordings during sensor calibration and flow studies. Note that the working liquid in our study has a Newtonian behaviour which differs from the behaviour of blood. However, for vessels with diameter larger than 1 mm it has been proven in former studies that blood behaviour can be approximated as Newtonian fluid.[22, 23] A newly developed PIV plasma with rheological and coagulation properties has been reported by Clauser et al.[24]. The viscosity used in our study is at the upper range of their study.

The MHVs are mounted at the sinus root in the centre of the transparent container (see Fig. 3(a)) such that in both cases the left-most leaflet is facing the left sinus in the symmetry-plane. For the Triflo TMHV the other leaflets face then the direction of the other sinus bulbs too, while for the SJM Regent the right leaflet is in the middle of both. For reference measurements, a thin-walled nozzle with circular orifice (outer diameter 25mm, inner diameter 24.6mm) was inserted into the aorta in the same plane as the MHV, which blocks the region of the SOV and generates a smooth core flow into the aorta. This configuration is studied at the same physiological flow rate for calibration of the sensors with simultaneous micro-PIV measurements.

A transparent inlet tube is placed upstream of the MHVs, which is connected to a pulse flow generator via a flexible pipe. A physiological pulsatile flow is generated with a heart rate of 70 beats per minute and a stroke volume of 80ml. To ensure also physiological conditions of the pressure across the valves, a constant pressure head of 80mmHg is applied to the open exit of the aorta by using a pressurized air cushion in the upper half of the tank above the liquid level.

**High-speed particle image velocimetry**

A continuous wave Argon-Ion laser (Raypower 5000, 5 W power at $\lambda$ = 532 nm, Dantec Danamics) is used as illumination source. The output laser beam is about 1.5 mm in diameter and is further expanded to a sheet illuminating the symmetry plane of the aorta. Full format recordings of the flow in the AAo are done with a high-speed camera (Phantom Miro 310, Ametek, 576×768 px$^2$ recording at 7200 fps) equipped with a lens (Tokima Macro $f$ = 100mm, F 2.8) showing a field of view of 57.6 x 76.8 mm² (see Fig. 4(b)). Zoomed-in recordings are carried out by an inverted telescope lens (Model K2/SC™, Infinity Photo-Optical Company, Boulder, USA), which yields a view field of view format of 4.0 x 6.4 mm² at 7200 fps. This equipment is used for the calibration of the micro-pillar sensors using simultaneous micro-PIV measurements.

The post-processing to capture the pillar tip motion and the flow field is done with an in-house Matlab code. For PIV, the images are masked within the pillar regions and processed with 2D cross-correlation of successive images to calculate the in-plane velocity vector fields. Processing is done in small interrogation windows with an iterative grid refinement method with final window size of 32x32 px$^2$. To illustrate the trajectory patterns of the jets in the flow the finite-time Lyapunov exponent (FTLE) is calculated from the velocity fields (LCS MATLAB Kit Version 1.0, Dabiri Lab, Stanford University) where Lagrangian coherent structures (LCSs) are shown by the largest values in the FTLE field, which has been reported as an efficient way to diagnose time-evolving patterns in dynamical systems[25]. The calculation is done backward in time with a time interval of 14 ms for 10 snapshots with an interval of 10 snapshots.

Another set of High-speed PIV measurements was done in the transparent inlet tube in the centre plane while the pulse generator was working in testing conditions. The axial velocity profile *u(r,t)* across the circular cross-section of the tube was determined from the PIV data and the flow rate *Q(t)* was then calculated from the integral:

$$Q(t) = \int_0^R u(r,t) 2\pi r \, dr \,, \qquad (1)$$

where *R* is the inner radius of the inlet tube at the location of the measured profile.

**Wall shear stress imaging**

First wall shear stress measurements using flexible micro-pillar sensors were documented in Brücker et al.[26] and the method is applied herein for the comparative measurements of the wall shear stress downstream of different heart valve prostheses. The static and dynamic response of these sensors illustrated in Fig. 4(a) is characterized for different pillar shapes and fluids in our previous studies.[27, 28] For accurate measurements of the WSS there exist several conditions which need to be fulfilled: [27, 28]

- To avoid that gravitational forces influence the pillar motion, the density of the flexible micro-pillar sensor should be of the same order of the density of the working liquid, which is the case for the micro-pillars used herein

- Viscous forces should dominate the drag on the pillar, which is when the local Reynolds-number, characterized by the small scale of the pillar, $Re_{Tip} = \rho U_{Tip} d/\mu$ is of order *O*(10) or less. This ensures also that no vortex shedding occurs. An estimation with a tip velocity of half of the peak velocity *U$_P$* yields a value of *Re$_{Tip}$* = 10.

- The height of the micro-pillar should be such that it is in the linear range of the velocity profile near the wall. This is often a compromise between sensitivity and visibility and therefore a first order approximation is often estimated in the form $u(y) = U_{Tip}\, y/l + O(\varepsilon)$ with the characteristic velocity ($U_{Tip}$) at the tip. This approximation is valid if the pillar height is considerably smaller than the boundary layer thickness. From the open nozzle configuration, the average boundary layer thickness in the flow cycle is estimated to about 5mm. Thus the pillar is situated only in the first 20% of the boundary layer.

- Bending is small compared to the length of the pillars. Therefore, wall-normal forces are only of second-order influence compared to the wall-parallel shear. It is herein always below 10% of the length of the pillars.

- The time scale of the sensor response should be small compared to the characteristic time-scale of the flow. An estimation of the smallest time-scales in the aorta can be derived from the Kolmogorov scale as done in [29]. If we assume a typical pressure drop $\Delta p = 1000\ Pa$ of MHVs over a segment of flow of length $L = D_A$, the dissipation rate per unit of mass can be estimated as

$$\varepsilon = \frac{\Delta p Q}{\rho V} \approx 6.67\ W \cdot kg^{-1} \approx 6.67\ m^2 \cdot s^{-3} \qquad (2)$$

where $Q = U_m \pi D_A^2/4$ is the flow rate and $V = L \pi D_A^2/4$ is the volume of the flow segment. The temporal and spatial Kolmogorov scales are then defined according to [29] by

$$\tau_K = \left(\frac{\nu}{\varepsilon}\right)^{1/2}, \eta_K = \left(\frac{\nu^3}{\varepsilon}\right)^{1/4} \qquad (3)$$

respectively, where $v = \mu/\rho$ is the kinematic viscosity. Thus, the Kolmogorov scales for the considered aortic flow in our study are estimated to be $\tau_K \approx 0.8\ ms$ and $\eta_K \approx 60\ \mu m$. The response time of the micro-pillars is tested in a step response test where we applied a force to one of the pillars and recorded the relaxation back to equilibrium in the same liquid environment as in the aorta flow studies. The dynamic response is that of a clamped cantilever beam which is critically damped and is represented by the equations for a damped second order oscillator [27, 28]. The sensor response time is calculated to $\tau_{95} = 5ms$ (the response time is defined as the time required to reach 95% of the final equilibrium position) and the response remains nearly constant up to frequencies of 200Hz. Our focus is on oscillations in WSS that are induced by coherent flow structures, which are typically more than one order of magnitude larger in spatial and temporal scales than compared to the Kolmogorov scales. Previous studies confirmed that the important characteristics of heart valve flows is the dominance of large-scale coherent structures over any random fluctuations due to turbulence.[7] For the micro-pillar sensors applied herein, it is further assumed that only the low-frequency range portion persists within the very near-wall region, which is associated with the near-wall dynamics of the coherent vortex structures.[26] These arguments above show that the response time of the sensors is sufficient to address our research questions.

Under the above given assumptions the micro-pillar tip bending is then direct proportional to the wall shear rate with a constant response up to frequencies of about 200Hz [27, 28]. The proportionality factor is obtained in a calibration test under the same pulsatile flow conditions in the aortic arch where the micro-pillars are recorded simultaneous with micro-PIV measurements, in which the tip's displacement along the wall is calibrated by the wall-normal gradient of the tangential velocity.

The micro-pillar sensors labelled from 1 to 12 are placed uniformly along the outer aortic wall in the AAo (see Fig. 3(b)) in the middle between both model halves. The symmetry plane is where the flow is documented most of the time to be 2D with only weak out-of-plane motion, also seen from the original particle recordings by the long residence time of the particles in the light-sheet plane. We expect in this region peak values in WSS in streamwise direction when the side-orifice jets emanating from the gap between the valve ring and leaflets impact with the aortic wall. Two exemplary sensor locations are selected for the presentations of the results, one in the upstream region at pillar #4 and one in the downstream region at pillar #11. The tip displacement of the micro-pillar sensors is detected with sub-pixel resolution (optical measurement uncertainty <0.1px, maximum tip displacement 10px). A second order low-pass filter with cut-off frequency of 200Hz is used to reduce measurement noise above the frequency response of the sensor.

The measured tip displacement relative to the situation with liquid at rest is converted into the wall shear rate by calibration with the simultaneous micro-PIV recording, carried out in the open circular orifice configuration. The sketch of micro-PIV data processing is shown in Fig. 4(b). Cross-correlation of successive images is performed with a window size of 240×64 px$^2$ and 75% overlap to determine the velocity profile near the wall. One exemplary snapshot of the profile is plotted in Fig. 4(c) and the slope of the linear curve-fit is further taken as the value of the instantaneous streamwise velocity gradient at the wall, i.e. the wall shear rate. The peak value in the time-series of the micro-PIV measurements is used to determine the proportionality constant, which remains constant also for other instants in the flow profile. The comparative plots of the wall shear rate distribution over the systolic cycle (see Fig. 4(d)) show that the time-resolved micro-pillar signal coincides well with the results of the micro-PIV measurement. At the end of systolic cycle, the downstream near wall flow field is quite complex and often not enough tracer particle remain in the interrogation window which is why the micro-PIV data fail in the late phase of valve closure. Note, that imaging of the WSS distribution can be done simultaneous for 4-8 micro-pillar sensors while for the micro-PIV measurements very high seeding densities are required to get good PIV results near the wall, which is often not possible due to strong reflections and particle migration to the core flow away from strong near-wall velocity gradients. This is why WSS measurements using the micro-pillar sensors are ideally suited for complex flows as expected in the valve studies.

**3. Results**

**Inlet flow condition and leaflet kinematics**

The inlet flow profile is shown in see Fig. 5(a), with a systolic heart beat duration of $T_{sys}$=386 ms characterised by acceleration flow phases from 0 to $0.2T_{sys}$, peak systole phase from $0.2T_{sys}$ to $0.4T_{sys}$, deceleration flow phases from $0.4T_{sys}$ to $0.9T_{sys}$, and valve closure phases from $0.9T_{sys}$ to $T_{sys}$. At the peak of the flow, the peak inlet velocity reaches a value of $U_P$ = 0.95 m·s$^{-1}$. The mean velocity averaged over the complete beat cycle amounts to $U_m$ = 0.19 m·s$^{-1}$. The non-dimensional physical parameters of the flow are represented by the Reynolds number, Strouhal number and Womersley number:

$$Re_P = \frac{\rho U_P D_A}{\mu}, St = \frac{D_A}{2} \frac{f}{(U_P - U_m)}, \alpha = \frac{D_A}{2}\sqrt{\frac{2\pi f \rho}{\mu}}, \quad (4)$$

where $f$ is the heart rate. It is obtained that $Re_P = 5340$, $St = 0.019$ and $\alpha = 16$. The critical peak Reynolds number is calculated according to [30] based on the measured Strouhal number and Womersley number according to $Re_P^c = 169\alpha^{0.83} St^{-0.27}$, from which we obtain $Re_P^c = 4925$. The supracritical Reynolds number is $Re_P - Re_P^c = 415$, which indicates that the flow is unstable[30]. Also, as the non-dimensional physical parameters of the flow are similar to that observed in in-vivo experiment of the healthy aorta performed among 30 volunteers by Stalder et al. [30], it indicates that the flow condition used in our study is representative for the natural flow situation.

The flow measurements are supplemented by instantaneous positions of both leaflets of BMHV and of TMHV, taken from high speed recordings of the tip motion of the leaflets. Note, that all three leaflets of TMHV open and close simultaneously and therefore one profile is representative for all other leaflets[31]. Comparative results of the leaflet motion measurements are given in Fig. 5(b). The leaflets of TMHV reach a maximum opening position at an angle of about 75° after 0.1$T_{sys}$, while the leaflets of BMHV reach the maximum angle of 85° at 0.2$T_{sys}$. The leaflets of BMHV are observed to have a different start in closing related to the orientation of the leaflets relative to the sinuses, i.e. the left leaflet facing the left sinus closes later at $0.8T_{sys}$ while the right leaflet starts closing much earlier at $0.6T_{sys}$.

**Flow field**

Analysis of the velocity field is done by calculating the out-of-plane component of the vorticity by central difference scheme and displaying those as color-coded iso-contours overlaid with the velocity vector field in Fig. 6, with a comparison of flow evolution for the BHMV compared to TMHV. Further analysis by the FTLE field shown in Fig. 7 illustrates the jet pattern generated by both MHVs. Regions of high FTLE values allow to recognize the boundaries of the jets. In addition, regions of high vorticity near the wall let recognize the development of the boundary layers along the walls.

*SJM Regent BHMV.* During the valve opening phase (see Fig. 6(a)), the leaflets' leading edges move from outside towards the center. A pair of vortices occur at the outer edges of the leaflets by shear layer separation, marked as outer starting vortices in Fig. 6(a). Flow separation at the edge of the valve housing causes the anti-clockwise vortex penetrating into the left sinus region, which is the so-called aortic sinus vortex (ASV). These vortices affect the near wall flow in the aorta while they travel further up into the AAo (see Fig. 6(b)). Later, regular rows of discrete vorticity structures occur downstream of the leaflets with alternating sign, resembling the form of von Karman-type vortex streets.

The boundary between fluid entering into the aorta from the fluid inside the aorta above the valve is presented high FTLE values, see in the region of the left sinus in Fig.

7(a) and (d). A wash-out flow driven by the ASV can be observed and the generation of the left side orifice jet is seen in Fig. 7(b). Downstream of the trailing edges of both leaflets the above discussed vortex streets appear as vertical columns with a zig-zag type pattern of the shear layers. At peak systole shown in Fig. 7(c), the SOJ has attached to the left wall of the aorta and affects the outer aortic wall.

*Triflo TMHV.* In the opening phase of the Triflo, the tips of the leaflet move from the center towards the sinuses, which causes a different flow field than in SJM. A leading inner starting vortex if followed by an outer starting vortex when the left leaflet reaches the maximum opening position. Both travel further downstream while in the inner part of the valve the central jet builds up, seen by rather uniform core flow in Fig. 6(e) and (f).

The more significant high FTLE value shown at the left sinus region in Fig. 7(d) indicates that there is a stronger wash-out flow compared to BMHV, which is supported by the displacement fluid generated by of the outwards rotating leaflet. The separated shear layers at the outer and inner edge of the leaflet are presented by the side orifice jet and central orifice jet separately. Thereafter, these two jets mix together and travel towards the centre of the aorta instead of impacting on the vessel wall at peak systole (see Fig. 7(e) and (f)).

Besides, the absolute value of out-of-plane vorticity field is integrated across the entire flow field and normalized by the open orifice nozzle configuration, which gives a rough quantitative indication how chaotic the flow is due to extra vorticity generated in the jets and wakes of the valve-ring and occluders. The normalized integration of BMHV is $E_{BMHV} = 1.93$ at $0.18 T_{sys}$, $E_{BMHV} = 1.49$ at $0.25 T_{sys}$ and $E_{BMHV} = 1.45$ at 0.4$T_{sys}$ over peak systole phase. Compared to BMHV, the normalized integration of TMHV leads to values of $E_{TMHV} = 1.94$ at 0.18$T_{sys}$, $E_{TMHV} = 1.34$ at 0.25$T_{sys}$ and $E_{TMHV} = 1.27$ at 0.4$T_{sys}$. This indicates that the flow field is more chaotic in BMHV over peak systole phase. Note that this is obtained only from a single plane in the flow and therefore the conclusion may not necessarily holds for the entire flow field. Though, as this plane is the symmetry plane of the flow crossing the centre of the shear layers it is a strong indicative for the entire flow.

**Wall shear rate**

Comparative profiles of wall shear rate in BMHV, TMHV and open circular orifice nozzle configurations are presented for the systolic cycle. The circular orifice configuration can be regarded as a reference flow in the arch at identical physiological flow conditions as for the MHVs, but without the influence of valve bodies. The profiles of wall shear rate measured by the micro-pillar sensors at upstream region and downstream region are plotted in Fig. 8(a) and (b), respectively.

During the valve opening phase, the flow acceleration generates a progressive rise of the wall shear rate with similar shape for all configurations. In this early phase, the displacement effect generated by the opening valves is similar for all configurations and the WSS is mainly influenced by bulk flow movement. The measured profiles resemble in this phase strongly those of the reference flow without a valve. Thereafter, BMHV shows a further increase of wall shear rate, which is linked to a stronger wash-out flow from the side orifice jet. At around 0.2$T_{sys}$, a drop is seen for both MHVs. This is caused by the passing-by of the counter-clockwise-rotating vortex structure of the inner starting vortex which temporarily induces a reduction of the near-wall flow velocity.

Shortly thereafter, a large positive peak of the wall shear rate appears around the peak systole in BMHV, which is caused by the impact of the strong side orifices jet-like flow travelling the outer aortic wall in the AAo. The peak wall shear rate reaches 2123 $s^{-1}$ in BMHV while the peak wall shear rate in TMHV configuration is only 910 $s^{-1}$, which is less than half of the BMHV and of similar magnitude as observed for the reference case.

After this peak, it shows characteristic waviness of wall shear rate profiles, which is due to the shear-layer roll-up at the region between the core flow and the aortic wall when global flow starts decelerating. It can also be found that the waviness is larger in BMHV compared with the TMHV, which means the rolling-up of the shear layer along the outer aortic wall in the AAo of BMHV is more intense. This phenomenon is more significant at further downstream region of BMHV (see Fig. 8(b)) from $0.35T_{sys}$ to $0.6T_{sys}$. Moreover, the peak wall shear rate of BMHV is 2046 $s^{-1}$, which is still twice the value of the peak wall shear rate of TMHV in this region, i.e. 1030 $s^{-1}$.

**4. Discussion and Conclusion**

Our detailed experimental results of the flow field generated by BMHV in the aortic arch can be compared to previous studies by Dasi et al.[8] and De Tullio et al.[14] at similar non-dimensional flow parameters and inlet flow profiles. As we use a more realistic curved geometry of the arch compared to the straight axisymmetric aorta of Dasi et al., we observed almost an asymmetric flow. The left leaflet of BMHV facing directly the sinus is delayed in closing due to the HV, while the two leaflets in the study of Dasi et al. show a synchronous motion when closing. The leaflet motion applied in the numerical study performed by De Tullio et al. coincides with our results, as well as the structure of the asymmetric flow field supporting our flow field results. This also means the curved geometry of the aortic arch is important to recover the impact of flow structures on the aortic walls, which is the reason for high peaks in WSS.

During valve opening, a strong wash-out flow at SOV is observed in TMHV. It indicates that, compared to BMHV, TMHV has a better performance on assisting blood transportation to coronary arteries, which helps with blood supply to the heart muscle. In addition, the zig-zag type flow pattern representing the vortex streets in the wake of the leaflets of BMHV is not observed in TMHV configuration. Rather, the latter shows a more homogeneous core flow with less small-scale structures generated in the shear layers. Besides, the side orifice jet of BMHV impacts on the outer wall of the AAo and causes peak WSS values with about three-fold magnitude compared to the reference flow at the same inlet flow profile. In comparison, the results of TMHV show that the impact of orifice jets generated by TMHV is much weaker as for the BMHV. A first small peak WSS appears at the same time for both MHVs and for the reference flow, which is induced by the initial wash-out flow. Shortly thereafter, the jet-like flow generated by BMHV impacts on the outer aortic wall, causing a steep increase to high peaks of WSS. As the side orifice jet of BMHV propagates earlier into the aorta than the central orifice jet, the WSS signal shows the maximum peak earlier in time than the TMHV. In comparison, the later peak in THMV is caused by the central jet impacting further downstream the aortic wall. This indicates that not only the magnitude of mechanical force, but also the impact regions are specific to the different valves designs.

The numerical simulation of the flow in the human aorta by Lantz et al.[32], based on MRI-acquired geometry and flow rate of a subject specific aorta, shows a similar peak wall shear rate of about 800 $s^{-1}$ in the downstream region for a natural valve compared to our open orifice nozzle configuration at similar pulsatile flow (reference flow). In addition, their comparative study of a rigid aorta with a flexible one using fluid-structure interaction (FSI) has proven that there is only a small difference in terms of the magnitude of the instantaneous WSS in the AAo. This shows that our WSS measurements provide reliable values although our model neglects flexible walls. When transferring the shear rates into stresses by multiplying with the dynamic viscosity, the mechanical forces along the aortic wall reach peak stresses of roughly 100 dyn·$cm^{-2}$ for BMHV and only half about that for TMHV compared to the reference case with about 40 dyn·$cm^{-2}$. These values are still about one order of magnitude larger than the WSS values reported by Meierhofer et al[16] obtained in an in-vivo MRI study of the flow in the AAo. However, it is not clear if their reported values are averages along the circumference of the walls. In general, the spatial resolution of MRI of about 1mm is not sufficient to capture the steep velocity gradients near the wall, therefore we assume a strong underestimation of the true WSS. However, while their absolute magnitude of WSS values might be much too low, the comparative effects observed between tricuspid and bicuspid aortic valves should represent the relative differences in a realistic manner. Their results showed a 15% higher WSS in the AAo for patients with bicuspid aortic valve caused by inborn connective tissue disease compared to the natural tricuspid aortic valves. They concluded that this increase is correlated with the often-observed dilation of the aorta in such patients. Transferring this conclusion to our comparative WSS measurements it indicates that TMHV may have the potential to largely reduce mechanical forces on the aortic walls compared to the standard mechanical prostheses having only two leaflets.

In summary, the main focus of our study is to investigate the mechanical load imposed by WSS along the vessel wall in the AAo when MHV prostheses are implanted. Our study describes new direct time-resolved WSS measurements with an array of calibrated sensors distributed along the curved aortic wall, by which we investigate the effects of MHVs on WSS distribution. The technique of micro-pillar WSS imaging is applied to achieve details of the shear stress working on the blood cells near the wall regions and the endothelial cells during a whole systolic cycle, which normally is very difficult to achieve. The results show that the peaks in WSS are located with the impact regions of the valve-specific orifice jets and differ largely with different valve designs.

**5. Acknowledgements**

This work is sponsored by BAE System and the Royal Academy of Engineering (RCSRF1617\4\11), both jointly funding the position of Professor Christoph Bruecker, which is gratefully acknowledged herein. The position of MSc Qianhui Li is supported by the German Research Foundation in the grant DFG 1494/32-1, which is also gratefully acknowledged. We thank the Helmholtz Institute of RWTH Aachen University for providing us the SJM Regent valve and Didier Lapier for the Triflo valve.

**6. Funding**

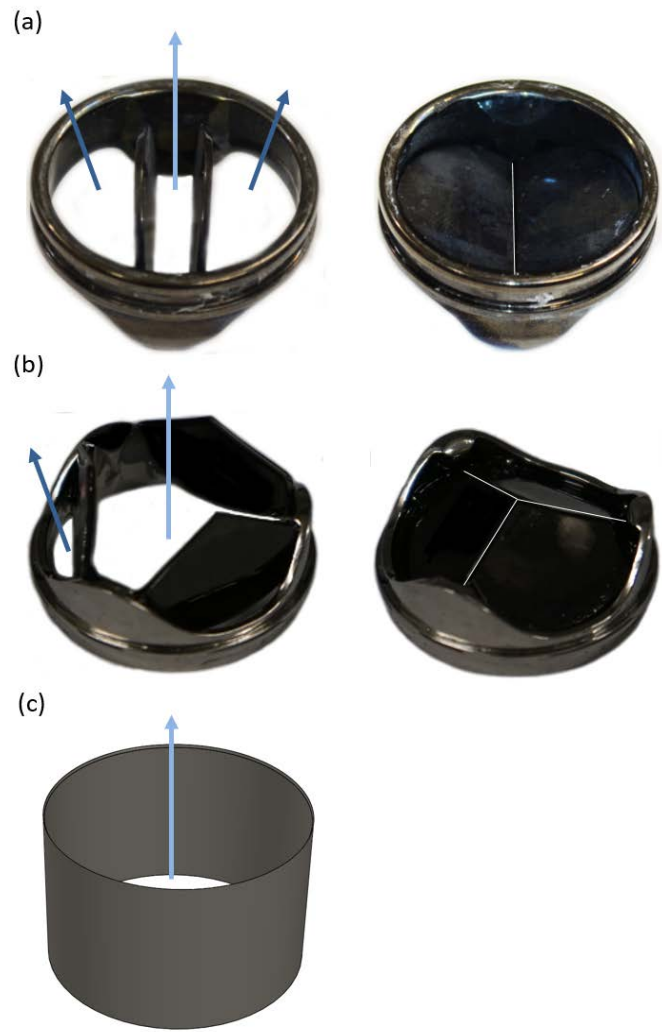

**Fig. 1.** (a) SJM Regent BMHV, with the leaflets in fully open position (opening angle: 85°) and in the closed position (closed angle 25°). (b) Triflo TMHV, with the leaflets in fully open position (opening angle: 75°) and in the closed position (closed angle 40°). (c) Thin-walled nozzle with circular orifice. Light blue arrow marks the centre orifice jet while dark blue arrow marks the side orifice jet.

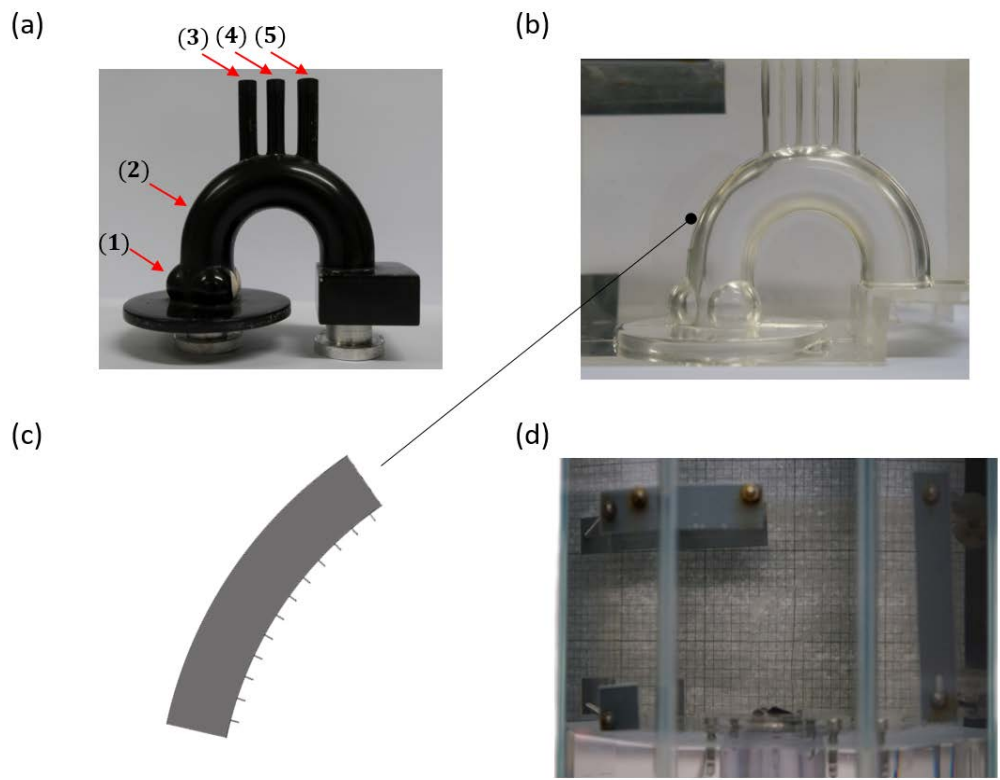

**Fig. 2.** Transparent silicone model of the aorta with the components to manufacture the model such as (a) the kernel of the aorta and (b) the final model assembled by the two symmetric halves. The foil with the pillars sketched in (c) is clamped between both halves. (d) Side view of the fluid basin with a millimetre paper in the back to demonstrate the refractive index matching.

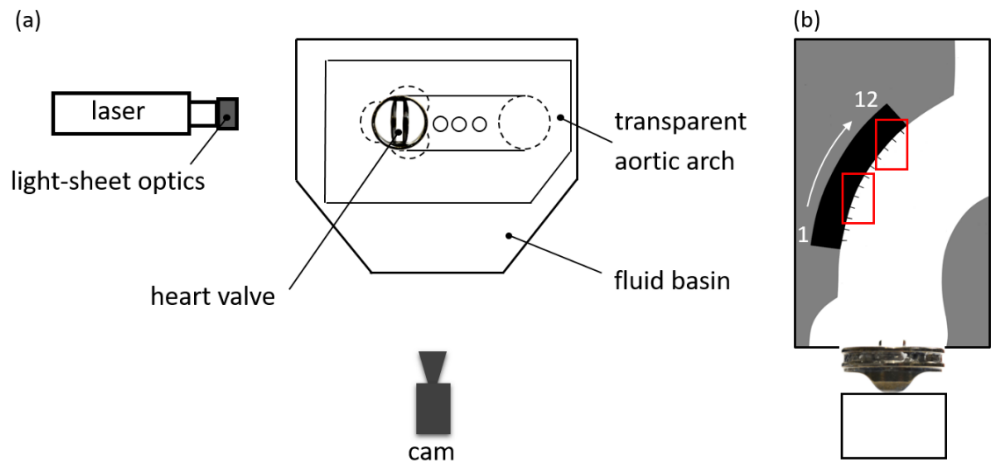

**Fig. 3.** (a) Top view of the experimental setup. (b) Frontal view of the measurement regions in the centre plane. Full AAo (the black rectangle downstream of MHV): 46x73.7 mm². Zoom-in region (red rectangle): 4.0 x 6.4 mm². Inlet flow measurement region (the black rectangle upstream of MHV): 25 x 11 mm². PDMS sheet with micro-pillars is shown in black.

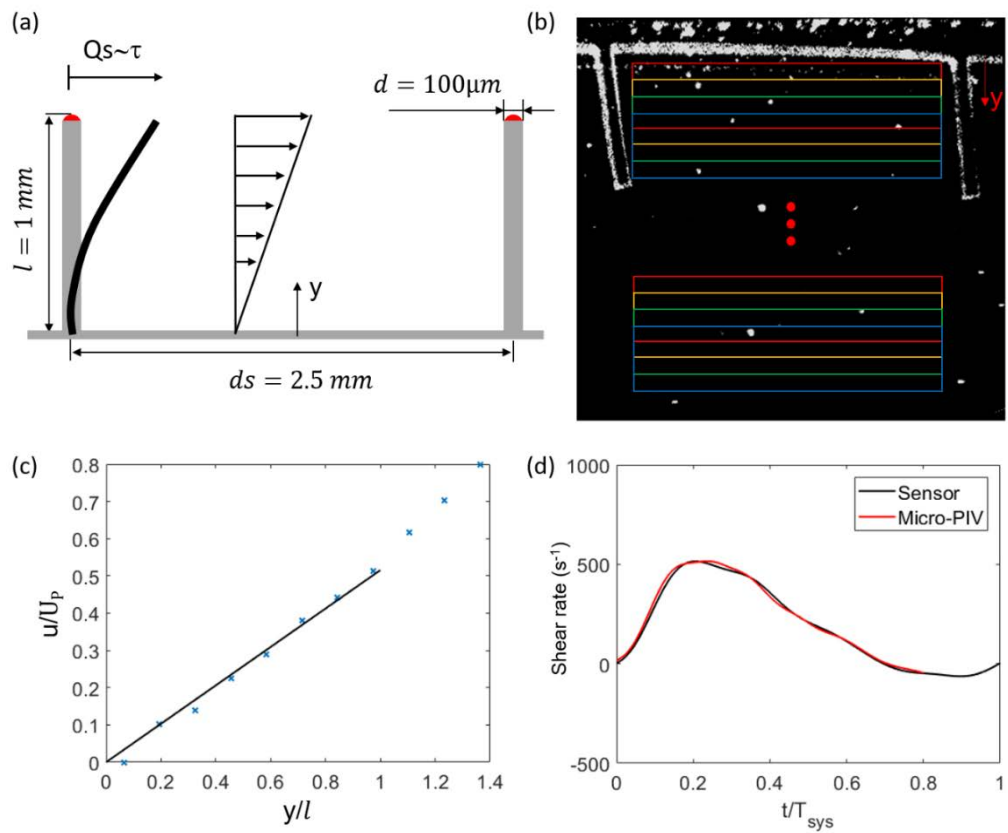

**Fig. 4.** (a) Parameters and the working principle of the micro-pillar sensors. (b) The sketch of the micro-PIV data processing. The cross-correlation window is presented by the coloured rectangles with 75% overlap of adjacent windows. (c) Plots of near wall velocity measured by the micro-PIV with blue marks and linear curve-fit. (d) Plot of the wall shear rate over the systolic cycle in reference flow situation.

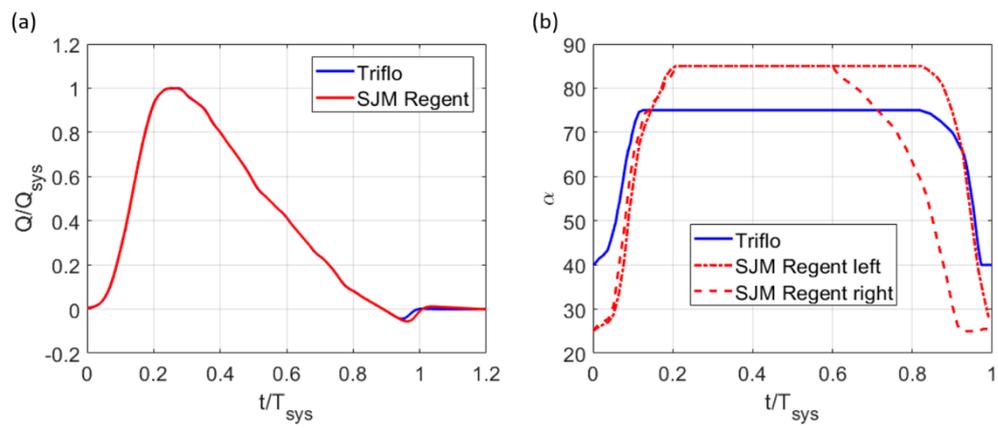

**Fig. 5.** (a) Inlet flow profile measured from high-speed PIV in the inlet tube and calculated using eq. (1), where $Q_{sys}$ is the peak flow rate over the systolic cycle. (b) Profile of the motion of the leaflet's angular position over the systolic cycle.

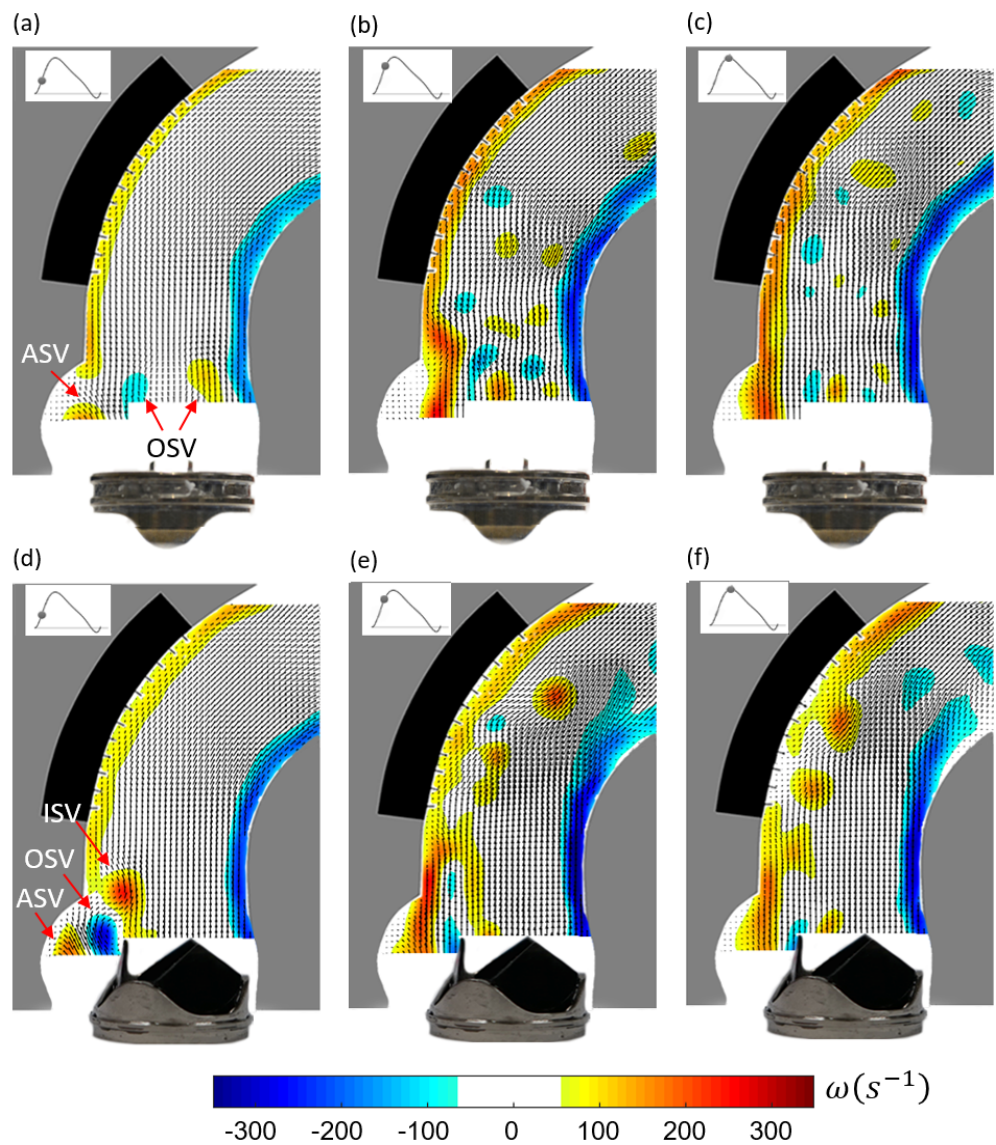

**Fig. 6.** Color-coded distribution of the out-of-plane component of the vorticity overlaid on snapshot vector fields at (a)(d) $t/T_{sys}$ = 0.1, (b)(e) $t/T_{sys}$ = 0.18, (c)(f) $t/T_{sys}$ = 0.25. OSV marks outer side (near the SOV) starting vortex. ISV marks inner side (near the centreline) starting vortex.

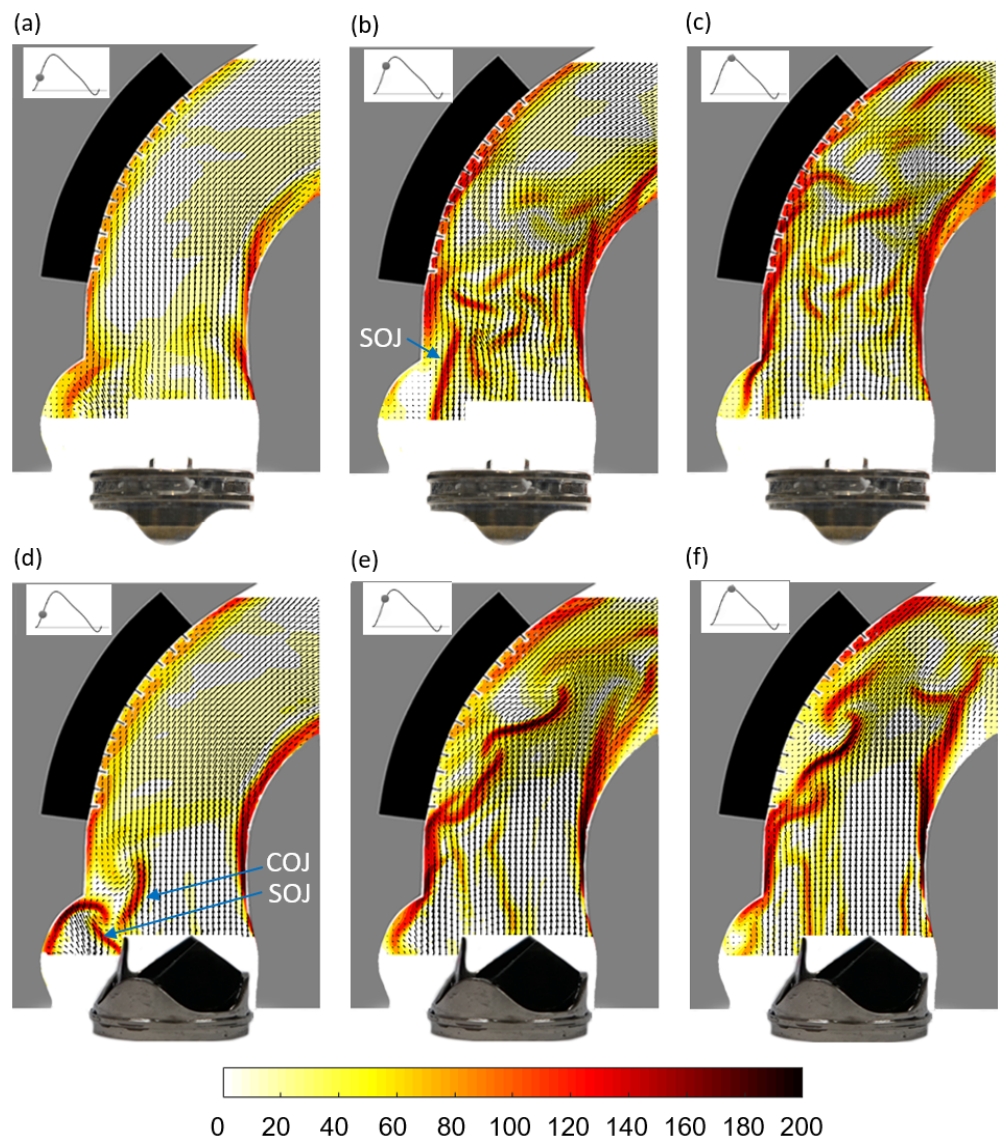

**Fig. 7.** Evolution of jet trajectories illustrated by the colored FTLE overlaid on snapshot vector fields at (a)(d) $t/T_{sys}$ = 0.1, (b)(e) $t/T_{sys}$ = 0.18, (c)(f) $t/T_{sys}$ = 0.25. SOJ marks side orifice jet. COJ marks centre orifice jet.

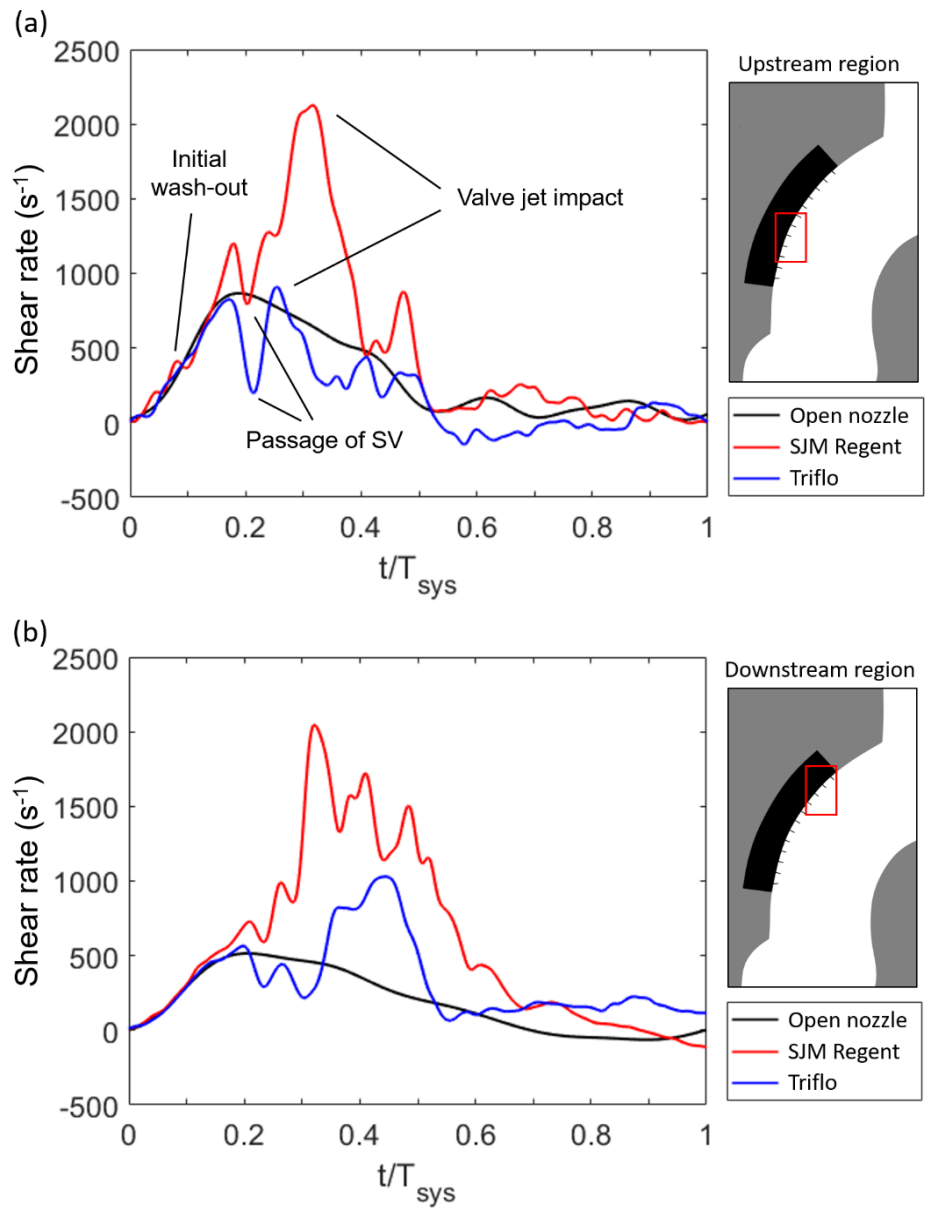

**Fig. 8.** Plots of wall shear rate on (a) upstream region and (b) downstream region of the outer wall of the AAo over the systolic cycle. Black line: open circular orifice nozzle, red line: SJM Regent BMHV, blue line: Triflo TMHV.